\documentclass[twocolumn,showpacs,superscriptaddress]{revtex4}
\usepackage{graphicx}
\usepackage{amsmath}
\usepackage{amsthm}
\usepackage{amssymb}
\usepackage{latexsym}
\usepackage{array}
\usepackage{float}
\usepackage{amsfonts}
\usepackage{mathrsfs}
\usepackage{verbatim}

\usepackage{color}

\begin{document}

\title{Quantifiable simulation of quantum computation beyond stochastic ensemble computation}

\author{Jeongho Bang}\thanks{Corresponding authors}
\affiliation{School of Computational Sciences, Korea Institute for Advanced Study, Seoul 02455, Korea}

\author{Junghee Ryu}
\affiliation{Centre for Quantum Technologies, National University of Singapore, 3 Science Drive 2, 117543 Singapore, Singapore}
\affiliation{Institute of Theoretical Physics and Astrophysics, Faculty of Mathematics, Physics and Informatics, University of Gda\'{n}sk, 80-308 Gda\'{n}sk, Poland}

\author{Chang-Woo Lee}
\affiliation{School of Computational Sciences, Korea Institute for Advanced Study, Seoul 02455, Korea}

\author{Ki Hyuk Yee}
\affiliation{School of Electrical and Computer Engineering, University of Seoul, Seoul 02504, South Korea}

\author{Jinhyoung Lee}\thanks{Corresponding authors}
\affiliation{Department of Physics, Hanyang University, Seoul 04763, Korea}

\author{Wonmin Son}\thanks{Corresponding authors}
\affiliation{Department of Physics, Sogang University, Seoul 04107, Korea}

\received{\today}

\begin{abstract}
In this study, a distinctive feature of quantum computation (QC) is characterized. To this end, a seemingly-powerful classical computing model, called ``stochastic ensemble machine (SEnM),''  is considered. The SEnM runs with an ensemble consisting of finite copies of a single probabilistic machine, hence is as powerful as a probabilistic Turing machine (PTM). Then the hypothesis---that is, {\em the SEnM can effectively simulate a general circuit model of QC}---is tested by introducing an information-theoretic inequality, named readout inequality. The inequality is satisfied by the SEnM and imposes a critical condition: if the hypothesis holds, the inequality should be satisfied by the probing model of QC. However, it is shown that the above hypothesis is not generally accepted with the inequality violation; namely, such a simulation necessarily fails, implying that PTM $\subseteq$ QC. 
\end{abstract}


\maketitle

\newcommand{\bra}[1]{\left<#1\right|}
\newcommand{\ket}[1]{\left|#1\right>}
\newcommand{\abs}[1]{\left|#1\right|}
\newcommand{\expt}[1]{\left<#1\right>}
\newcommand{\braket}[2]{\left<{#1}|{#2}\right>}
\newcommand{\commt}[2]{\left[{#1},{#2}\right]}

\newcommand{\tr}[1]{\mbox{Tr}{#1}}

\newcommand{\note}[1]{\textcolor{blue}{#1}}

\section{Introduction}

By relating computation to physics, Feynman asked, ``can (quantum) physics be simulated by (classical) computers?'' and gave the negative answer \cite{Feynman86}: It {\em might be} impossible to efficiently simulate a quantum process on a probabilistic Turing machine (PTM) \footnote{Similarly, around this time, Benioff also argued that quantum computation may be at least as powerful as classical computation by showing how quantum physics can simulate the computational process of a classical reversible Turing machine \cite{Benioff82}.}. Then, a quantum computer that could simulate any quantum system was conjectured by Feynman. Following Feynman's original arguments, Deutsch developed a physically realizable model of a quantum computer, i.e., a quantum Turing machine (QTM), a quantum analogue of the PTM \cite{Deutsch85}. After that, a formal argument that a QTM can be more powerful than a PTM was provided \cite{Bernstein93,Simon97}, and quantum computation (QC) has been investigated intently and more deeply with the advent of celebrated quantum algorithms \cite{Deutsch92,Grover97,Shor97}. Today, it is widely believed that QC solves hard problems much faster. However, skepticism toward QC still exists since the identification of a clear border between classical and quantum computations is still obscure \cite{Bennett97,Adleman97,Aaronson09}. This would arise without ruling out the potential of any classical probabilistic computation model that is believed to imitate the QC \cite{Khrennikov08,Paler13,Ionescu14}.

With these open problems in mind, here we attempt to find a dissimilar aspect between classical versus quantum computations (without identifying the computational complexity \cite{papadimitriou03}). For this purpose, we consider a seemingly-powerful classical computing machine, called ``stochastic ensemble machine (SEnM),'' which runs with an {\em ensemble} consisting of a large (even infinite, in principle) number of single probabilistic machines (for example, PTMs). Such SEnM can cover non-Markov chain computation. Then, we immediately ask, ``Can SEnM simualte QC?'' Our approach for seeking the answer is to test the following hypothesis: there exists an SEnM that effectively simulates a specific model of QC---we consider the quantum circuit model in this study. Here, by the simulation we mean that the SEnM is able to reproduce quantum transition probabilities between all possible intermediate steps of our testing QC model. The test is carried out on the basis of an information-theoretic temporal inequality, called ``readout inequality,'' which imposes the critical condition: if the above hypothesis holds, the inequality should also be satisfied by our testing QC model. However, a violation of this inequality can be observed, whereas it can never occur by the SEnM. Such a discrepancy tells that the above hypothesis is generally not accepted and such a simulation necessarily fails.

\section{QC versus SEnM}

We start with a brief description of QC. Conventionally, QC runs as follows. First, an input $\ket{\psi_0}$ is initialized, and then, we place it into a kernel operation (e.g., a quantum Fourier transformation or several iterations of a nontrivial transformation), which consists of the sequences of fundamental unitary (gates) operations (hereafter, denoted by $\hat{C}_\text{comp}$). Finally, the solution information is extracted from the output state \footnote{There is, of course, a scheme for QC, so-called measurement-based QC \cite{Raussendorf01,Nielsen05}. However, such a scheme is not of our current interest, at least in this work.}. We decompose $\hat{C}_\text{comp}$ into the number of computation steps, say $L$, such that 
\begin{eqnarray}
\hat{C}_\text{comp} = \hat{U}_L \cdots \hat{U}_{2}\hat{U}_{1}, 
\label{eq:C_comp}
\end{eqnarray}
where $\hat{U}_j$ ($j=1,2,\ldots,L$) are the unitary transformations, realized as the possible transition maps in the QTM \cite{Yao93}. Thus, the action of $\hat{C}_\text{comp}$ is described as $\ket{\psi_{0}} \rightarrow{\hat{U}_{1}} \ket{\psi_{1}} \rightarrow{\hat{U}_{2}} \cdots \rightarrow{\hat{U}_{L}} \ket{\psi_{L}}$, where $\ket{\psi_j} = \sum_{m_j} \omega_{m_j} \ket{m_j}$, and $m_j \in {\cal M}$ denotes the readout symbol (${\cal M}$ is a set of symbols). The quantum probability $P_Q(m_j|\psi_j)$ of measuring $m_j$ at $j$ is then defined as \cite{Deutsch85}
\begin{eqnarray}
P_Q(m_j|\psi_j) = \abs{\omega_{m_j}}^2 = \abs{\braket{m_j}{\psi_j}}^2.
\label{eq:q_pro}
\end{eqnarray}
We note that the probabilities $P_Q(m_j|\psi_j)$ are abstract mathematical quantities and are not characterized in the middle of the QC in general. However, they are to be evaluated for our specific purpose, as described later.

Here, it is worth noting that there is a novel QC model, called duality QC, which is originally proposed to exploit the wave-particle duality so that it can also use the linear combinations of $U_j$'s \cite{Long06,Gudder07,Long11}. The duality QC appears to be more general than the typical QC model \cite{Long11} and offers more flexibility in quantum algorithm design \cite{Wei16,Wei17}. However, it will be sufficient for our purpose to consider the (typical) QTM that runs the products of $\hat{U}_j$'s as in Eq.~(\ref{eq:C_comp}); i.e., the superiority of the QTM over its classical counterparts, {\em if any}, is directly generalized to that of the duality QC model.

We then describe a seemingly-powerful classical computing machine---an SEnM---that runs in parallel with an ensemble of {\em indistinguishable} single-machine (e.g., PTM) copies. Each copy consists of two components---a finite processor and an infinite memory (or `tape')---of which only a finite portion is ever used, as in a conventional Turing machine \footnote{An additional element---a ``cursor''---is necessary to address the currently scanned memory location. However, it will be omitted throughout this paper, as our main conclusion remains valid (see Refs.~\cite{Deutsch85,Galindo02} for a more detailed summary of the Turing machine).}. Each copy is assumed to be definitely in one of the following possible states: for the $k$-th copy, its state $s_k \in {\cal S}$, where ${\cal S}$ is a finite set of possible states. The state collection of all copies is denoted by ${\mathbf s}_j = \left(s_1, s_2, \cdots\right)_j$ at any step $j$ of the computation. Here, let $m(s_k) \in {\cal M}$ be the readout (or measurement) symbol of the $k$-th copy $s_k$. Then, the probability $P(m_j|{\mathbf s}_j)$ of reading the symbol $m_j$ at step $j$ can be defined as
\begin{eqnarray}
P(m_j | {\mathbf s}_j) = \lim_{N \rightarrow \infty} \frac{1}{N} \sum_{k=1}^{N} \delta_{m(s_k), m_j},
\label{eq:p_mo}
\end{eqnarray}
and it characterizes the collection ${\mathbf s}_j$. Here, $\delta_{m,m'}$ is the Kronecker delta, and $\sum_{m_j} P(m_j | {\mathbf s}_j) = 1$. Note that these probabilities $P(m_j | {\mathbf s}_j)$ can be characterized in the computation, however, the readout symbol $m(s_k)$ of a specific $k$-th copy cannot be identified due to its indistinguishability.

\begin{figure*}
\centering
\includegraphics[width=0.9\textwidth]{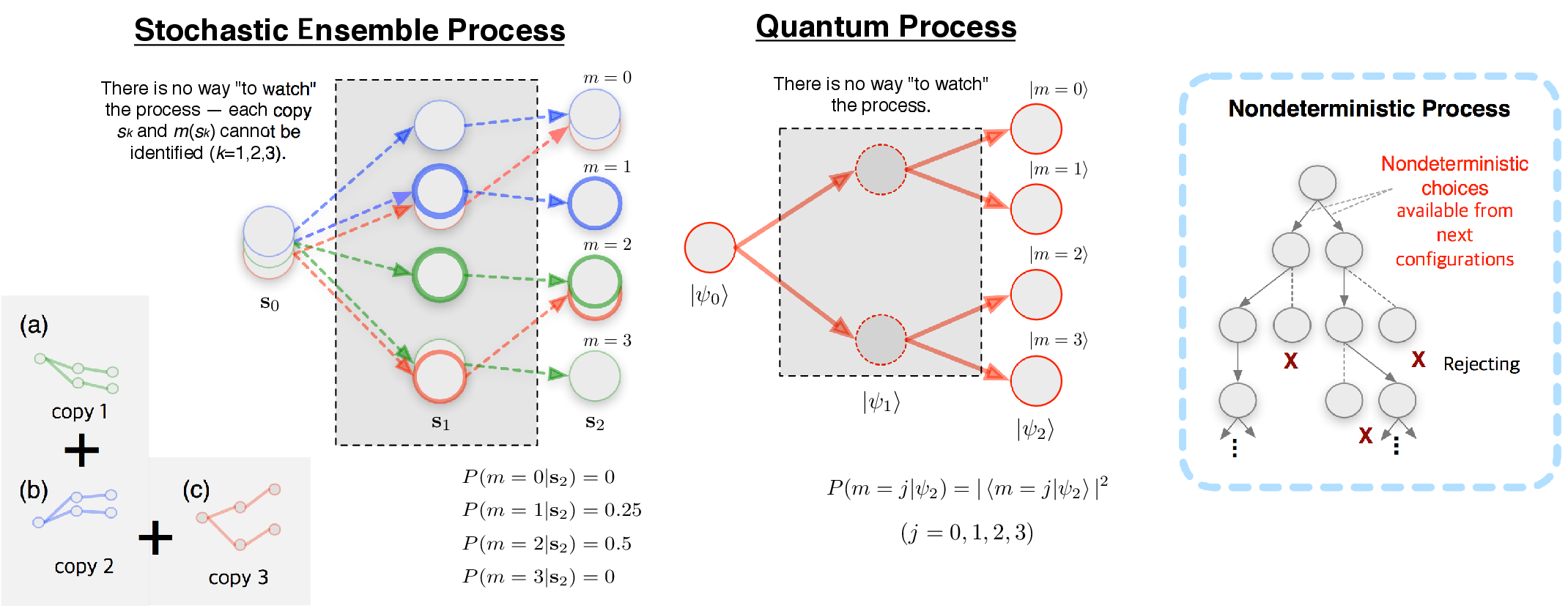}
\caption{\label{fig:E_comp} (color online) A schematic of the SEnM computation. For simplicity, we consider only three copies. Each copy has its own potential computation routes that reach their branches toward a local output area [as depicted in (a), (b), and (c)]. Here, the computation route (i.e., ${\bf s}_j \to {\bf s}_{j+1}$) are constructed by all of these copies---the computation is stochastic in this sense---however, the SEnM {\em cannot identify} the changes of the copies (i.e., $m_j(s_k) \to m_{j+1}(s_k)$). In a QC, it is possible to explore all computation routes simultaneously and we also {\em cannot identify} the computation. Here, we also illustrate the computation of a ``nondeterministic Turing machine,'' which can even cast and/or reject the computation routes concurrently \cite{papadimitriou03}. Of course, such an extraordinary ability is far from being realized.}
\end{figure*}

Each copy is allowed to have its {\em own} computation routes, which may potentially reach the desired solution. The SEnM, however, executes a {\em stochastic computation} in which the probability of reaching the targets is given by the sum of the probabilities along the possible routes of the readout symbols of every $k$-copy. However, particular routes of the readout symbols (e.g., $m_{j-1}(s_k) \to m_j(s_k)$) cannot be followed due to the indistinguishability of the single-machine copies. Therefore, a single step of the computation, a transition ${\mathbf s}_{j-1} \to {\mathbf s}_j$, is stochastic and is specified by the conditional probabilities (see Fig.~\ref{fig:E_comp}). Here, a set of such conditional probabilities corresponds to a table of the computation instructions in a Turing machine \cite{Galindo02}. More explicitly, a conditional probability $C({\mathbf s}_1 | {\mathbf s}_0)$ defines the first computational step in which the ensemble is in ${\mathbf s}_1$ from a given initial ${\mathbf s}_0$. In general, the computation at $j$-th step is defined by the conditional probability $C({\mathbf s}_j | {\mathbf s}_{j-1},...,{\mathbf s}_0)$, conditioned to the state collections ${\mathbf s}_{j'}$ ($j' < j$), which were obtained in the earlier steps (except the initial ${\mathbf s}_0$). Thus, we have a joint probability up to the $j$-th step, given in terms of the conditional probabilities $C$ by
\begin{eqnarray}
P({\mathbf s}_0,...,{\mathbf s}_j) = C({\mathbf s}_j|{\mathbf s}_{j-1}, ..., {\mathbf s}_0) \cdots C({\mathbf s}_1|{\mathbf s}_0) P({\mathbf s}_0),
\label{eq:jpdsc}
\end{eqnarray}
where $P({\mathbf s}_0)$ is the initial distribution of $\mathbf{s}_0$. Such an SEnM is clearly beyond a Markov chain computation \cite{Zukowski14}, even though it is not possible to identify the states of the individual copies in the computation. If we can design an SEnM that is able to identify each copy and its state, it is possible to follow the different copy-states concurrently in the computation. However, it is trivial to consider such an (imaginary) SEnM, since it is quite similar to a non-deterministic Turing machine (NTM) (see Fig.~\ref{fig:E_comp}). Note that the NTM is believed to be superior to the QTM; for example, NP-complete problems are polynomially solvable by the NTM but not by the QTM (for more details, see Ref.~\cite{papadimitriou03}).

\section{Hypothesis and Readout Inequality}

Now, we shall test if there exists an SEnM model that can simulate QC. To this end, we define the conditional probability of {\em reading out} a symbol $m_j$, given $m_{j-1}$ at every $j$-th step, as
\begin{eqnarray}
P(m_j|m_{j-1}) = \frac{P(m_{j-1}, m_j)}{P(m_{j-1})}.
\label{eq:nbcp}
\end{eqnarray}
The pairwise joint probabilities $P(m_{j-1}, m_j)$ are given as the marginals of the {\em grand joint probability of the readout symbols}:
\begin{eqnarray}
P(m_{0}, ..., m_{j}) = \sum_{\{ {\mathbf s}_j \}} P({\mathbf s}_0,...,{\mathbf s}_j) \prod_{l=0}^j P(m_l|{\mathbf s}_l).
\label{eq:grand_jp}
\end{eqnarray}
Then, we specify our hypothesis as follows.
\begin{itemize}
\item[{\bf H}] {\em There exists an SEnM such that it provides the conditional probabilities $P(m_{j}|m_{j-1})$ equal to $P_Q(m_j|m_{j-1})$ at each $j$-th step:
\begin{eqnarray}
P(m_j|m_{j-1}) = P_Q(m_j|m_{j-1})
\label{eq:correspond1}
\end{eqnarray}
for all $ j = 1,\ldots, L$, and then,
$P(m_L|m_0)$ for obtaining the final computation outcome $m_L$ given the initial $m_0$ is equal to the quantum probability $P_Q(m_L|m_0)$:
\begin{eqnarray}
P(m_L|m_0) = P_Q(m_L|m_0).
\label{eq:correspond}
\end{eqnarray}}
\end{itemize}
Here, it is easily inferred that the inverse of the hypothesis {\bf H} is true; namely, there exists a QTM that statistically simulates the SEnM computation. For example, one can drive his/her QTMs in a statistical way, such that $\bra{m_j}\hat{\rho}_j\ket{m_j}$ is equal to $P(m_j | \mathbf{s}_j)$, where $\hat{\rho}_j$ is a {\em statistical mixture} of the symbol states decohered from $\ket{\psi_j}$.

To test this hypothesis {\bf H}, we use an information-theoretic inequality called the readout inequality as
\begin{eqnarray}
H(M_{L}|M_{0}) \le \sum_{j=1}^{L} H(M_{j}|M_{j-1}),
\label{eq:ineq}
\end{eqnarray}
which is derived by the validity of the grand joint probability $P(m_0, \ldots, m_L)$ in Eq.~(\ref{eq:grand_jp}) (for details of the derivation, see Appendix~\ref{appendix:A}). Here, $H(M_{l'}|M_{l})= -\sum_{m_l,m_{l'}} P(m_l, m_{l'}) \log_2{P(m_{l'}|m_l)}$ are the conditional Shannon entropies ($l' > l \in [0,L]$) \cite{Hamming80,Dembo91}, where $P(m_l , m_{l'})$ are the marginals given from the grand joint probability $P(m_0, \ldots, m_L)$ in Eq.~(\ref{eq:grand_jp}), and the conditional probabilities $P(m_{l'}|m_l)$ are defined as Eq.~(\ref{eq:nbcp}). This readout inequality imposes the critical condition: if the hypothesis {\bf H} holds, the readout inequality is also obeyed by QC with the SEnM simulating it.

The test of {\bf H} using the readout inequality in Eq.~(\ref{eq:ineq}) is suggestive of those that have been discussed in physical models of macro-realism and/or local-realism. In a realistic theory, the measurement results are regarded as {\em a-priori} properties independent of observation and they are carried by a ``hidden'' variable, which is often called reality. In addition to this, the macro-realism model assumes that any measurements performed at an instant of time has no influence on the subsequent dynamics, which is called non-invasive measurability. Similarly, the local-realism model assumes that the results obtained at a place are independent of any measurements performed at space-like separated place, which is called locality. These constraints can be tested by Leggett-Garg inequality and Bell inequality, respectively; namely, the inequalities should be satisfied with the aforementioned constraints. Quantum theory does not agree with both models violating the inequalities (for more details, see Refs.~\cite{Kofler08,Kofler13}). Adopting the view of these physical models (if one may particularly wish), an SEnM implicitly assumes the reality, i.e., the identity of the stochastic computation with $\{\mathbf{s}_0, \ldots, \mathbf{s}_L\}$. This condition is equivalent to the existence of the probability distribution $P(\mathbf{s}_0, \ldots, \mathbf{s}_L)$ defined in Eq.~(\ref{eq:jpdsc})~\cite{Fine82}. It also has the non-invasive measurability, namely that reading the symbol $m_l$ at the $l$-th step does not have any effect on the state collection ${\mathbf s}_l$ itself and on the readouts at the subsequent steps. Thus, in such a framework, we can predict that a violation of the readout inequality is forbidden for any computation in which the above two assumptions are implied.

Here, it should also be highlighted that the quantum violations of the Leggett-Garg and/or Bell inequalities can be simulated by a particular model where a certain amount of information can be communicated between locally or temporally separated testing points~\cite{Toner03, Brierley15}. Nevertheless, we clarify that the violation of the readout inequality in Eq.~(\ref{eq:ineq}) can never be observed in our SEnM, because there is no communication of the readout symbols $m$ of the single-machine copies; i.e., $m$ is given for the current ensemble state ${\mathbf s}$, as defined in Eq.~(\ref{eq:p_mo}). Note however that the SEnM can utilize the every information about the past history of the ensemble in the stochastic computation, as described in the previous section [also, refer to Eq.~(\ref{eq:jpdsc})].

\section{QC Violation of Readout Inequality}

From now on, we investigate whether the hypothesis {\bf H} is true in general. As the readout inequality is obeyed by {\em every} SEnM computation, the most general and direct way is to examine whether or not the inequality in Eq.~(\ref{eq:ineq}) can be violated by QC. Thus, we first consider a specific QC scheme, the so-called quantum amplitude amplification (QAA), where the iterative runs of a given unitary $\hat{G}_g$ enhance the probability amplitude (not the probability itself) of the target $\ket{\tau}$ among the number of candidates (say, $N$) \cite{Brassard00,Biham00}. Here, it is useful to know that, for a certain condition, $\hat{G}_g$ is described as a rotation of the angle $\Theta$ on the plane of the target $\ket{\tau}$ and its orthogonal state $\ket{\tau^\perp}$ as (for details, see~\ref{appendix:B})
\begin{eqnarray}
\hat{G}_g = -e^{i \Theta \hat{\sigma}_x},
\label{eq:G_g}
\end{eqnarray}
where $\Theta \le 2\sqrt{P_Q(m_0=\tau|\psi_0)}$, and $\hat{\sigma}_x=\ket{\tau}\bra{\tau^\perp} + \ket{\tau^\perp}\bra{\tau}$. Note that $\Theta$ is supposed to be small because the initial probability $P_Q(m_0=\tau|\psi_0)$ is vanishingly small with a large $N$. The total number of iterations $n_c$ is given as 
\begin{eqnarray}
n_c = \text{CI}\left[ \frac{1}{\Theta}\left({\frac{\pi}{2} - \sqrt{P_Q(m_0=\tau|\psi_0)}}\right)\right],
\end{eqnarray}
where $\text{CI}[x]$ is the closest integer to $x$. Here, if $P_Q(m_0=\tau|\psi_0) = \frac{1}{N}$, we can obtain a quadratic speedup with $n_c = O(\frac{\pi}{4}\sqrt{N})$ compared to $O(N)$ of the classical strategy \cite{Grover97,Zalka99}.

Then, we show that a violation of the inequality can occur in QAA. First, we define $\hat{C}_\text{comp}$ by cutting the entire process of QAA; namely, we only consider a finite number of iterations $n_d$ ($< n_c$) such that 
\begin{eqnarray}
\hat{C}_\text{comp} = \hat{U}_L \cdots \hat{U}_2 \hat{U}_1 = \hat{G}_g^{n_d}
\end{eqnarray}
by letting $\hat{U}_j = \hat{G}_g^{{n_d}/{L}}$ ($\forall j$). Here, $n_d$ and $L$ are initially chosen so that ${n_d}/{L}$ becomes an integer number. To proceed, we define the measurement $\hat{M}_j = \hat{\Pi}_{m_j \neq \tau} - \hat{\Pi}_{m_j = \tau}$, where $\hat{\Pi}_{m_j=\tau}=\ket{\tau}\bra{\tau}$, and $\hat{\Pi}_{m_j\neq\tau}=\hat{\openone} - \ket{\tau}\bra{\tau}$. We then have
\begin{eqnarray}
P_Q(m_j) &=& \tr{(\hat{\rho}_j \hat{\Pi}_{m_j})}, \nonumber \\
P_Q(m_{j}|m_{j-1}) &=& \abs{\bra{m_j}{\hat{G}_g^{{n_d}/{L}}}\ket{m_{j-1}}}^2, 
\end{eqnarray}
where $\hat{\rho}_j = \ket{\psi_j}\bra{\psi_j}=\left.\hat{G}_g^{({n_d}/{L}) j}\right.\hat{\rho}_0\left.\hat{G}_g^{({n_d}/{L})j}\right.^\dagger$. Again, we note that these measurements are performed only for the purpose of our inequality test. Using Eq.~(\ref{eq:G_g}) and by simple calculations, the inequality in Eq.~(\ref{eq:ineq}) is rewritten as
\begin{eqnarray}
h\left({n_d}\Theta\right) \le L h\left(\frac{n_d}{L}\Theta\right),
\label{eq:ineq_q}
\end{eqnarray}
where $h(x) = -\left(\cos^2{x}\right) \log_2{\left(\cos^2{x}\right)} - \left(\sin^2{x}\right)\log_2{\left(\sin^2{x}\right)}$. Here, let us define the quantity 
\begin{eqnarray}
D = L h\left(\frac{n_d}{L}\Theta\right) - h\left({n_d}\Theta\right)
\end{eqnarray}
so that a negative value of $D$ indicates a violation and its degree. In such settings, let $n_d \to \frac{\pi}{4\Theta}$ by taking $n_d$ to be $\frac{1}{2}n_c$. Note that, if $n_c$ is odd, $n_d = \lfloor \frac{1}{2}n_c \rfloor$, where $\lfloor x \rfloor$ is the floor of $x$, the largest integer less than or equal to $x$. Then, $h({n_d}\Theta)$ approaches the theoretical maximum; i.e., $h({n_d}\Theta) \to h(\frac{\pi}{4})=1$. In contrast, $L h\left(\frac{n_d}{L}\Theta\right) \to L h\left(\frac{\pi}{4L}\right)$, which can still be less than $1$. This directly results in a violation; namely, 
\begin{eqnarray}
D = L h\left(\frac{\pi}{4L}\right)-1 < 0
\end{eqnarray}
for an appropriately chosen $L$. Here, the maximum violation is given as 
\begin{eqnarray}
D_\text{min}=\text{CI}\left[\frac{\pi}{4\Theta}\right] h\left(\Theta\right)-h\left(\text{CI}\left[\frac{\pi}{4\Theta}\right]\Theta\right)
\end{eqnarray}
with $L = n_d = \text{CI}\left[\frac{\pi}{4\Theta}\right]$. We particularly infer that $D_\text{min} \to -1$ when $\Theta \to 0$, or equivalently, $N \to \infty$ (for more detailed analyses, see~\ref{appendix:C}).

Therefore, it is clear that, in general, QC does not agree with a legitimate grand joint probability distribution $P(m_0, \ldots, m_L)$ [as in Eq.~(\ref{eq:grand_jp})] constructed from the conditional probabilities $P_Q(m_j|m_{j-1})$. This points toward the invalidity of the hypothesis ${\mathbf H}$. 

\section{Summary and Discussion}

To summarize, we have studied a distinguishable feature of a specific model of QC by considering a powerful classical computing machine called the SEnM. Our main question was ``can SEnM simulate QC?'' and the answer was ``generally, no.'' To obtain such an answer, we specified the following hypothesis {\bf H}: there exists an SEnM that statistically simulates QC. The test of this hypothesis was carried out with the readout inequality. By establishing the subtly connected link between the physical assumptions implicitly involved in {\bf H}, we explicitly showed that there is a discrepancy between a QC and the SEnM's QC simulation; QC can violate the readout inequality, even though the two computations should satisfy the readout inequality if the hypothesis {\bf H} holds. This discrepancy directly indicates that {\bf H} is not generally accepted. And further, this result implies that it is generally not possible to prepare a classical setting which produces the desired output with the same number of QC steps, particularly in a quantum circuit model. Thus, noting that SEnM $\supseteq$ PTM, we can conclude that QC is characterized beyond the PTM.

\section*{Acknowledgments}

We are grateful to Marcin Paw\l{}owski, Marek \.{Z}ukowski, and Ryszard Horodecki for helpful discussions and comments. JB and JL acknowledge the financial support of the Basic Science Research Program through the National Research Foundation of Korea (NRF) funded by the Ministry of Science, ICT \& Future Planning (No. 2014R1A2A1A10050117) and the MSIP (Ministry of Science, ICT and Future Planning), Korea, under the ITRC (Information Technology Research Center) support program (IITP-2017-2015-0-00385) supervised by the IITP (Institute for Information \& Communications Technology Promotion). JR acknowledges the TEAM project of FNP, EU project BRISQ2, and the National Research Foundation and Ministry of Education in Singapore. KHY acknowledges the support of IITP funded by the Korea government (MSIT) (No. 2017-0-00266, Gravitational effects on the free space quantum key distribution for satellite communication). JR also acknowledge the National Research Foundation, Prime Minister's Office, Singapore and the Ministry of Education, Singapore under the Research Centre of Excellence programme and Singapore Ministry of Education Academic Research Fund Tier 3 (Grant No. MOE2012-T3-1-009). WS acknowledges the ICT R\&D program of the MSIP/IITP (No. 2014-044-014-002) and the National Research Council of Science and Technology (NST) (Grant No. CAP-15-08-KRISS). WS also acknowledges the support by KIAS through their open KIAS program.

\appendix

\section{Derivation of the inequality in Eq.~(\ref{eq:ineq})}\label{appendix:A} 

We derive the inequality in Eq.~(\ref{eq:ineq}) by assuming the validity of the grand joint probability $P(\{m_j\})$ for $j=0,1,\ldots,L$ in Eq.~(\ref{eq:grand_jp}). First, consider the joint Shannon entropy
\begin{eqnarray}
H(\{M_j\}) = - \sum_{\mathbf{m}} P(\{m_j\}) \log_2 P(\{m_j\}), 
\end{eqnarray}
where $\sum_\mathbf{m}$ stands for $\sum_{m_{L}}\cdots\sum_{m_{1}}\sum_{m_{0}}$, and the base of the logarithm function is chosen to be $2$. By the chain rule for the Shannon entropy, we may factor the joint entropy $H(\{M_j\})$ as
\begin{eqnarray}
H(\{M_j\}) &=& H(M_{L} | M_{0}, \ldots, M_{L-1}) \nonumber \\
                && + H(M_{L-1} | M_{0}, \ldots, M_{L-2}) \nonumber \\
                && + \cdots + H(M_{1} | M_{0}) + H(M_{0}), 
\label{eq:jH}
\end{eqnarray}
where $H(M_{l'}|M_l)$ ($l' > l \in [0, L]$) is the conditional Shannon entropy defined by
\begin{eqnarray}
H(M_{l'}|M_l) = -\sum_{m_l,m_{l'}} P(m_l, m_{l'}) \log_2{P(m_{l'}|m_l)}. 
\end{eqnarray}
Here, the probabilities $P(m_l, m_{l'})$ are the marginals from $P(\{m_j\})$, and $P(m_{l'}|m_l)$ are given from Eq.~(\ref{eq:nbcp}).
By employing the information inequalities \cite{Hamming80,Dembo91}, 
\begin{eqnarray}
&& H(M_{j} | M_{0}, \ldots, M_{j-1}) \le H(M_{j} | M_{j-1}), \nonumber \\
&& H(M_{0}, M_{j}) \le H(M_{0}, M_{1}, \ldots, M_{j}), 
\end{eqnarray}
we arrive at the information-theoretic inequality
\begin{eqnarray}
H(M_{L}|M_{0}) \le \sum_{j=1}^{L} H(M_{j}|M_{j-1}).
\end{eqnarray}

\section{Quantum amplitude amplification}\label{appendix:B} 

In general, the operation $\hat{G}_g$ consists of two reflections as \cite{Brassard00,Biham00}
\begin{eqnarray}
\hat{G}_g = -\hat{I}_{Q(m)}\hat{I}_{\phi}. 
\end{eqnarray}
Here, $\hat{I}_{\phi}$ is the quantum Householder reflection defined by $\hat{I}_{\phi} = \hat{\openone} - (1-e^{i \phi}) \ket{\psi_0}\bra{\psi_0}$, where $\hat{\openone}$ is the identity operation. The phase $\phi$ is generally given from $0$ to $\pi$. Another reflection $\hat{I}_{Q(m)}$ is the quantum oracle operation defined by $\hat{I}_{Q(m)} = \sum_{m}e^{i Q(m)}\ket{m}\bra{m}$, where $Q(m)$, called the ``query function,'' is defined such that $Q(\tau)=\varphi_\tau \in (0, \pi]$ and $Q(m)=0$ for all $m \neq \tau$. Here, we also note that $\hat{I}_{Q(m)} = \hat{\openone} - (1-e^{i \varphi_\tau})\ket{\tau}\bra{\tau}$.

For further analysis, we let 
\begin{eqnarray}
\ket{\tau^\perp} = \frac{\ket{\psi_0} - \sqrt{P_Q(m_0=\tau|\psi_0)}\ket{\tau}}{\sqrt{1-P_Q(m_0=\tau|\psi_0)}},
\end{eqnarray}
where it is assumed that $\ket{\tau^\perp}$ absorbs the relative phase. Then, $\hat{G}_g$ is given as a $2 \times 2$ unitary matrix on the plane of $\ket{\tau}$ and $\ket{\tau^\perp}$, assuming $O\left(P_Q(m_0=\tau|\psi_0)\right) \rightarrow 0$, as
\begin{widetext}
\begin{eqnarray}
\hat{G}_g=
\left(\begin{array}{cc}
-e^{i \varphi_\tau} & \left(1-e^{i \phi}\right)\sqrt{P_Q(m_0=\tau|\psi_0)} \\
\left(1-e^{i \phi}\right)e^{i \varphi_\tau}\sqrt{P_Q(m_0=\tau|\psi_0)} & -e^{i \phi}
\end{array}\right).
\label{eq:a_matrix}
\end{eqnarray}
\end{widetext}
For a large-scale (i.e., $N \gg 1$) problem, such an assumption is reasonable because $P_Q(m_0=\tau|\psi_0)$ would be vanishingly small. We then write $\hat{G}_g$ explicitly, neglecting the global phase factor $e^{-i\frac{\phi + \varphi_\tau}{2}}$, as follows:
\begin{eqnarray}
\hat{G}_g = \hat{R}_{\varphi_\tau/2} \left[ \left(\begin{array}{cc} e^{i \frac{\xi}{2}} &  \\  & e^{-i \frac{\xi}{2}} \end{array}\right) + i \Theta \hat{\sigma}_x \right]  \hat{R}_{\varphi_\tau/2}^\dagger,
\label{eq:a_decom}
\end{eqnarray}
where
\begin{eqnarray}
\hat{R}_{\varphi_\tau/2} = \left(\begin{array}{cc} 1 & 0 \\ 0 & -e^{i \frac{\varphi_\tau}{2}} \end{array}\right),
\end{eqnarray}
\begin{eqnarray}
\xi &=& \phi - \varphi_\tau, \nonumber \\
\Theta &=& 2\sqrt{P_Q(m_0=\tau|\psi_0)}\sin{\frac{\phi}{2}},
\end{eqnarray}
and $\hat{\sigma}_x = \left(\small{\begin{array}{cc}  0 & 1 \\  1 & 0 \end{array}}\right)$ in the bases of $\{\ket{\tau}, \ket{\tau^\perp}\}$. We rewrite $\hat{G}_g^{2^m}$ in Eq.~(\ref{eq:a_decom}) as
\begin{eqnarray}
\hat{G}_g^{2^m} = \hat{R}_{\varphi_\tau/2} \hat{H}_{\Theta,\xi}^{2^m}  \hat{R}_{\varphi_\tau/2}^\dagger, 
\label{eq:a2m}
\end{eqnarray}
where $\hat{H}_{\Theta,\xi}$ is defined by
\begin{eqnarray}
\hat{H}_{\Theta,\xi} = \left(\begin{array}{cc} e^{i \frac{\xi}{2}} &  \\  & e^{-i \frac{\xi}{2}} \end{array}\right) + i \Theta \hat{\sigma}_x.
\label{eq:h}
\end{eqnarray}
The operation $\hat{H}_{\Theta,\xi}^{2^l}$ is evaluated, for $l=1,2,\ldots,m$, as
\begin{eqnarray}
\hat{H}_{\Theta,\xi}^{2^1} &=& \left(\begin{array}{cc} e^{i 2 \frac{\xi}{2}} &  \\  & e^{-i 2 \frac{\xi}{2}} \end{array}\right) + i 2 \cos\left(2^{-1}\xi\right)\Theta \hat{\sigma}_x, \nonumber \\
\hat{H}_{\Theta,\xi}^{2^2} &=& \left(\begin{array}{cc} e^{i 2^2 \frac{\xi}{2}} &  \\  & e^{-i 2^2 \frac{\xi}{2}} \end{array}\right) + i 2^2 \cos\left(2^{-1}\xi\right) \cos\left(2^{0}\xi\right) \Theta \hat{\sigma}_x, \nonumber \\
&\vdots& \nonumber \\
\hat{H}_{\Theta,\xi}^{2^m} &=& \left(\begin{array}{cc} e^{i 2^m \frac{\xi}{2}} &  \\  & e^{-i 2^m \frac{\xi}{2}} \end{array}\right) + i 2^m \Gamma \Theta \hat{\sigma}_x,
\label{eq:h2m}
\end{eqnarray}
where
\begin{eqnarray}
\Gamma = \prod_{j=0}^{m-1}\cos{\left(2^j \frac{\xi}{2}\right)}=\frac{1}{2^m} \frac{\sin{\left(2^m \frac{\xi}{2}\right)}}{\sin{\frac{\xi}{2}}}.
\label{eq:Gamma_h}
\end{eqnarray}
Here, if we assume that the phase factor $e^{-i \frac{\varphi_\tau}{2}}$ is also absorbed into $\ket{\tau^\perp}$, we can ignore $\hat{R}_{\varphi_\tau/2}$ and $\hat{R}_{\varphi_\tau/2}^\dagger$ in Eq.~(\ref{eq:a2m}); thus, we have $\hat{G}_g^{2^m}=\hat{H}_{\Theta,\xi}^{2^m}$.

Then, by using Eqs.~(\ref{eq:h})--(\ref{eq:Gamma_h}), we can express $\hat{G}_g^{2^m}$ without a loss in generality as
\begin{eqnarray}
\hat{G}_g^{2^m} = -e^{-i {2^m} \Gamma \Theta \hat{\sigma}_x} - \left(\begin{array}{cc} 1-e^{-i 2^m\frac{\xi}{2}} &   \\  & 1-e^{i 2^m\frac{\xi}{2}} \end{array}\right),
\label{eq:a2m_2}
\end{eqnarray}
where the first term $-e^{ -i {2^m} \, \Gamma \Theta \, \hat{\sigma}_x }$ is the main process of amplitude amplification, and the last term represents the inaccuracy or error. Here, we note that the condition $\xi \rightarrow 0$, known as the ``phase-matching condition,'' should be satisfied to successfully complete amplitude amplification without error (see Refs.~\cite{Long99,Biham00} for the issues). Namely, as $\Gamma \rightarrow 1$ when $\xi \rightarrow 0$ [with the property of the sine function in Eq.~(\ref{eq:Gamma_h})], it is straightforward that
\begin{eqnarray}
\hat{G}_g^{2^m} = -e^{-i 2^m \Theta \hat{\sigma}_x}.
\label{eq:aop_rot2m}
\end{eqnarray}

\section{Detailed analyses of the inequality violations in QAA}\label{appendix:C} 

\begin{figure}[t]
\centering
\includegraphics[width=0.23\textwidth]{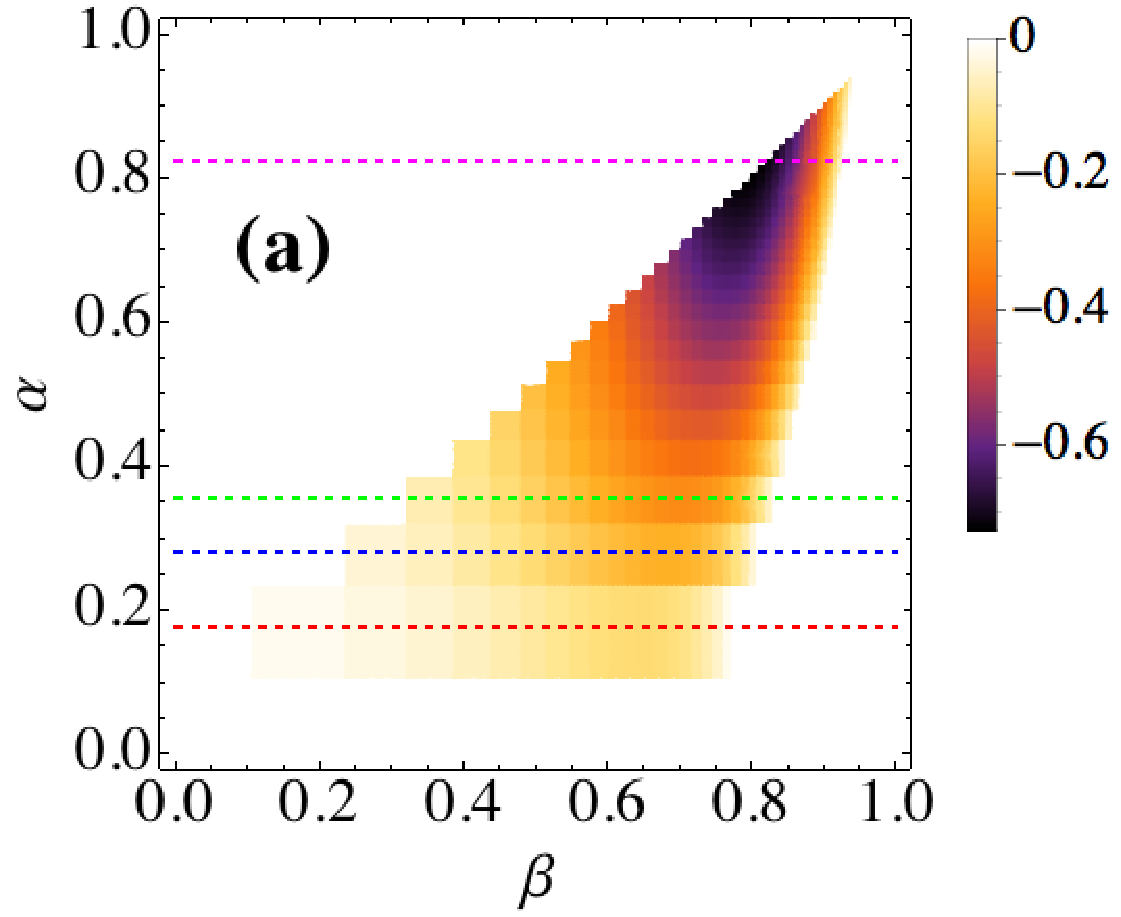}
\includegraphics[width=0.23\textwidth]{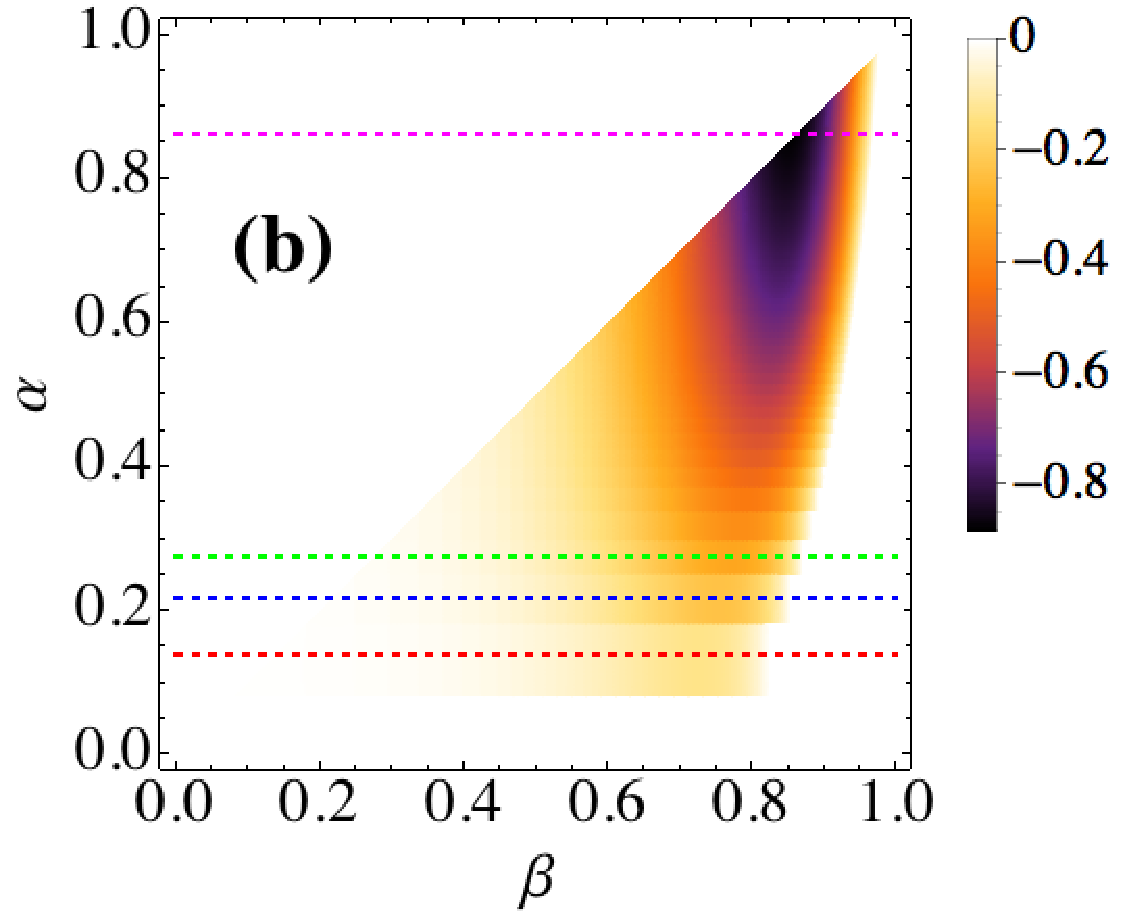}
\\
\includegraphics[width=0.23\textwidth]{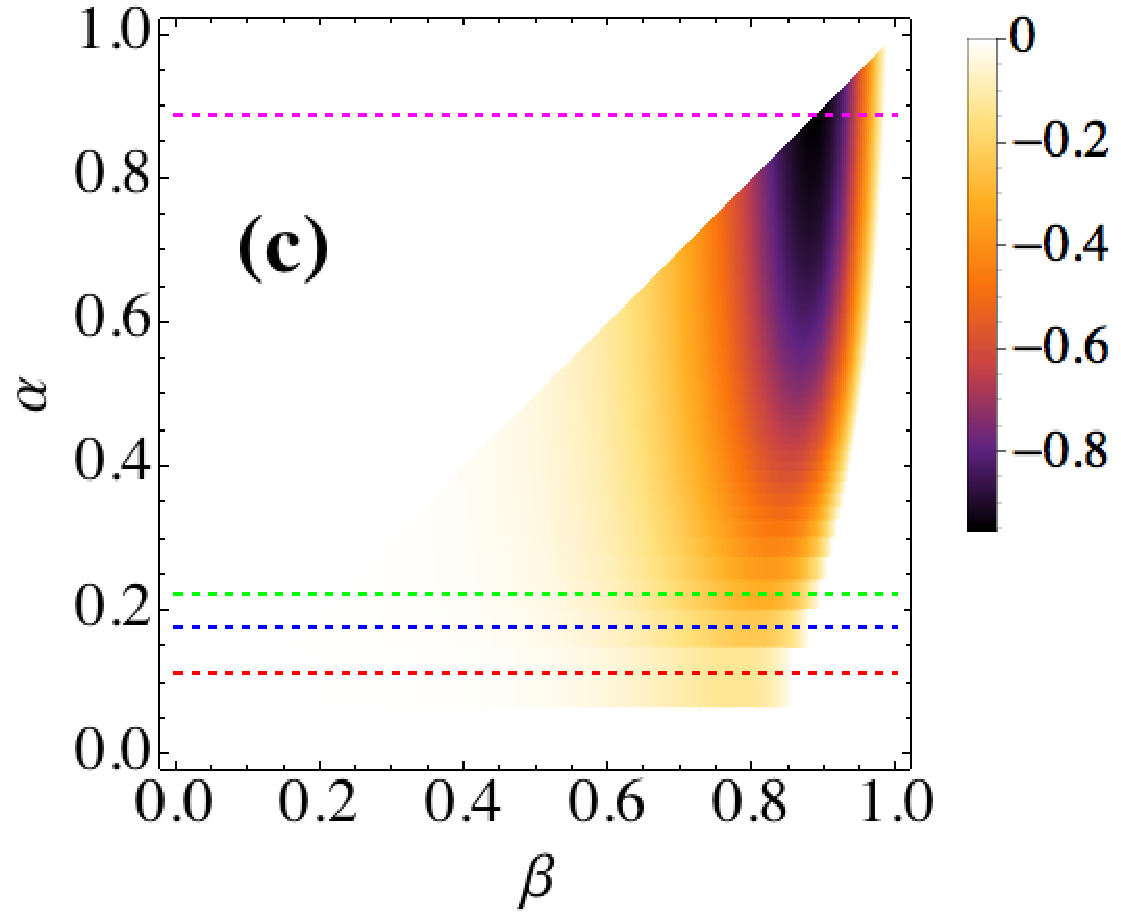}
\includegraphics[width=0.23\textwidth]{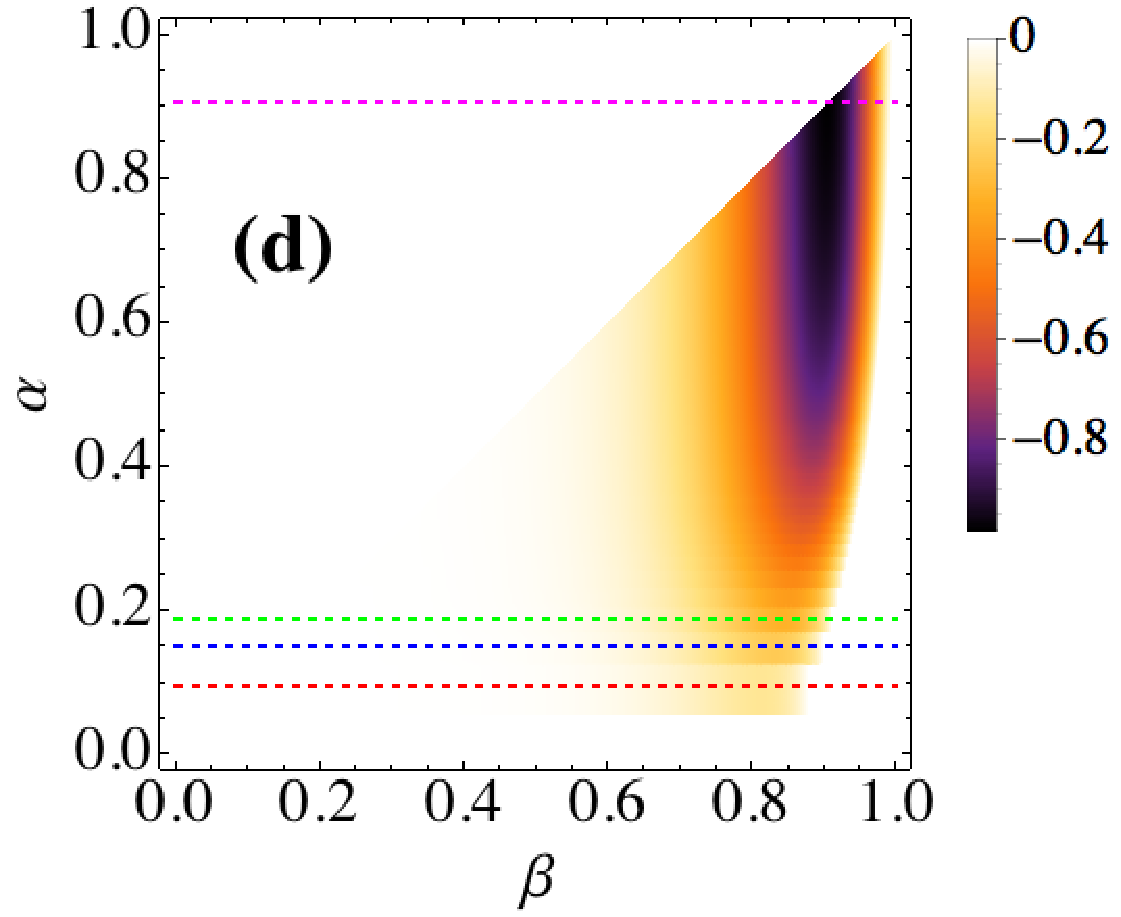}
\caption{(color online) We depict the inequality violations (i.e., the negative values of $D$) in the possible region of a two-dimensional plane ($\beta$, $\alpha$). Here, we consider four cases: (a) $\Theta=\sqrt{1/10^3}$, (b) $\sqrt{1/10^4}$, (c) $\sqrt{1/10^5}$, and (d) $\sqrt{1/10^6}$. In each graph, we also draw the lines for $\alpha$ corresponding to $L=2$ (red dashed), $L=3$ (blue dashed), $L=4$ (green dashed), and $L = \text{CI}(\frac{\pi}{4\Theta})$ (magenta dashed).}
\label{grp:violations_dp}
\end{figure}

\begin{figure}[t]
\centering
\includegraphics[width=0.23\textwidth]{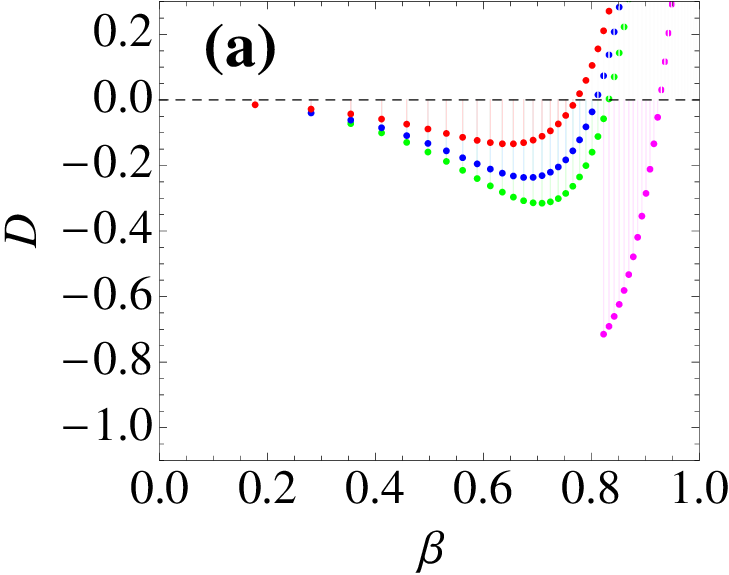}
\includegraphics[width=0.23\textwidth]{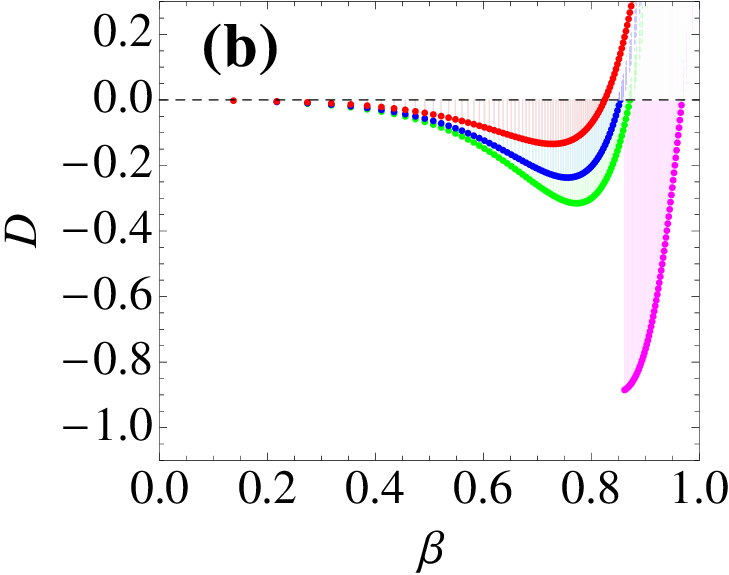}
\\
\includegraphics[width=0.23\textwidth]{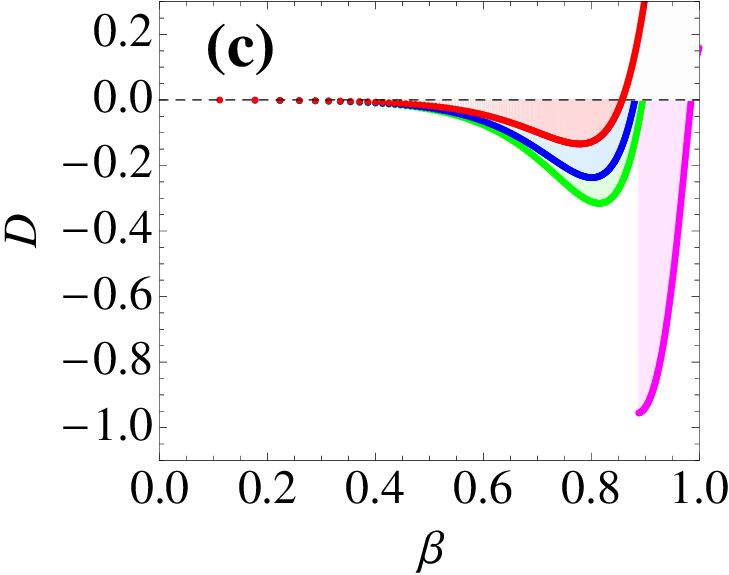}
\includegraphics[width=0.23\textwidth]{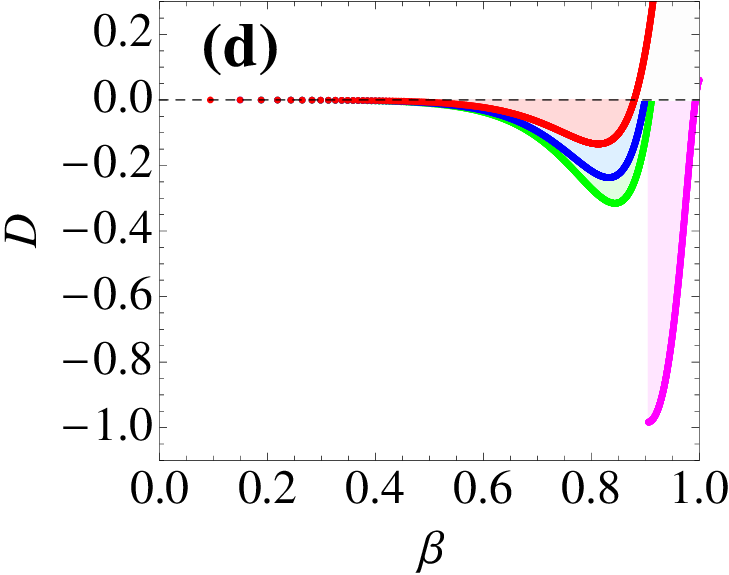}
\caption{(color online) Graphs of $D$ versus $\beta$ are drawn for $L=2$ (red points), $3$ (blue points), $4$ (green points), and $\text{CI}(\frac{\pi}{4\Theta})$ (magenta points). Each graph corresponds to that in Fig.~\ref{grp:violations_dp}: (a) $\Theta=\sqrt{1/10^3}$, (b) $\sqrt{1/10^4}$, (c) $\sqrt{1/10^5}$, and (d) $\sqrt{1/10^6}$. An inequality violation ($D < 0$) is observed even for a relatively small experimental scale (i.e., $L=2,3,4$), but the violations would be conspicuous when the experiment scale $L$ becomes large.}
\label{grp:violations_L}
\end{figure}

To see the violation more explicitly, we first define two parameters, $\alpha$ and $\beta$, related to the factors $L$, $n_d$, and $n_c$:
\begin{eqnarray}
\alpha = \log_{n_c}{L}, ~\text{and}~ \beta = \log_{n_c}{n_d},
\end{eqnarray}
where $0 \le \alpha \le \beta \le 1$, which is imposed by the condition $1 \le L \le n_d \le n_c$. Another parameter for the inequality test is $\Theta$, which is given in terms of $n_c$, or equivalently, the problem scale $N$. Note that $\Theta$ is assumed to be very small for very large $N$ in a practical application. For example, if we consider the original Grover's search algorithm, we have $\Theta=2\sqrt{1/N}$ for $\phi=\varphi_\tau=\pi$. In this circumstance, we show the graphs of the inequality violations ($D < 0$) in the possible region of the two-dimensional space of $\beta$ and $\alpha$, i.e., $\alpha \ge 0$, $\beta \ge 0$, and $\alpha \le \beta \le 1$ (see Fig.~\ref{grp:violations_dp}). Here, we consider four cases of $\Theta=\sqrt{1/10^{\gamma}}$ with (a) $\gamma=3$, (b) $\gamma=4$, (c) $\gamma=5$, and (d) $\gamma=6$. In each graph, we draw dashed lines for $\alpha$ corresponding to $L=2$ (red), $3$ (blue), $4$ (green), and $\text{CI}(\frac{\pi}{4\Theta})$ (magenta). Note that the maximum violation is observed on the line of $\text{CI}(\frac{\pi}{4\Theta})$. It is also seen that the violations are conspicuous when $L$ becomes large.

\begin{figure}[t]
\centering
\includegraphics[width=0.42\textwidth]{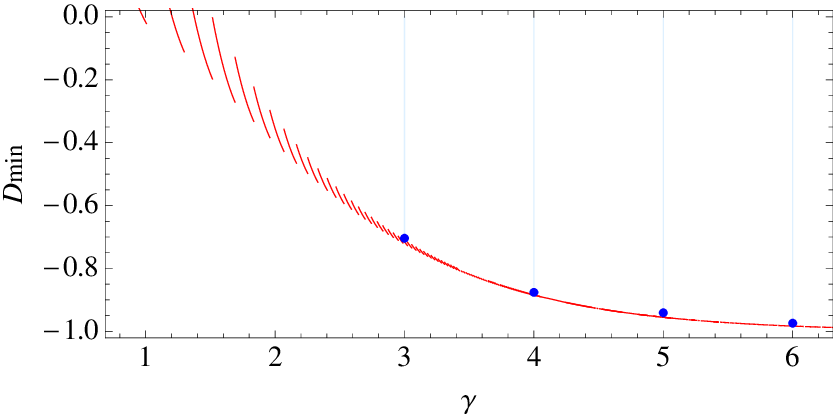}
\caption{(color online) We plot the simulation data (blue points) of the maximum violation, i.e., $D_\text{min}$, for $\gamma = 3,4,5,6$. The data are very well matched to the theoretical values (red lines) given by Eq.~(\ref{eq:Dmin}).}
\label{grp:Dmin}
\end{figure}

In Fig.~\ref{grp:violations_L}, we additionally present the values of $D$ along the lines of $L=2,3,4,\text{CI}(\frac{\pi}{4\Theta})$ drawn in Fig.~\ref{grp:violations_dp} by increasing $L$ from $2$ to $n_d$ (see the main text). Here, we note that our inequality test would be realized in an experiment for the reasonable problem scale $\Theta=\sqrt{10^{-\gamma}}$ and for a value of $L$ that is not too large because the size of the measurement settings becomes larger as $L$ increases. However, in order to see the maximum violation, i.e., $D_\text{min}$ (particularly, close to the lower bound $-1$), it is necessary to increase the measurement settings for a large problem scale $N$. In Fig.~\ref{grp:Dmin}, we plot the data of the maximum violation $D_\text{min}$ (blue points) obtained from Monte Carlo simulations compared with the theoretical values (red lines) given by
\begin{eqnarray}
D_\text{min}=\text{CI}\left(\frac{\pi}{4\Theta}\right) h\left(\Theta\right)-h\left(\text{CI}\left(\frac{\pi}{4\Theta}\right)\Theta\right).
\label{eq:Dmin}
\end{eqnarray}
The simulations are performed with the setting $L = n_d = \text{CI}(\frac{\pi}{4\Theta})$ to obtain the maximum violation by increasing $\gamma$. However, to reduce the computing flops, the simulations are performed on the two-dimensional space of $\ket{\tau}$ and $\ket{\tau^\perp}$ by employing Eq.~(\ref{eq:a_matrix}). The conditional and marginal probabilities for characterizing the conditional entropies are evaluated by averaging $10^4$ outcomes of the measurements at each step. Each point is made by averaging $1000$ trials of inequality tests. It is seen that the data are match the theoretical values well. The detailed values are listed in Tab.~\ref{tab:data_sim}.

\begin{table}[h]
\centering
\tabcolsep=0.2in
\begin{tabular}{c|c|c}
\hline\hline
$\gamma$ & $D_\text{min}$ (Simul.)  & $D_\text{min}$ (Theor.)  \\
\hline
$3$ & $\simeq -0.703 \pm 0.013$ & $\simeq -0.714$ \\
$4$ & $\simeq -0.878 \pm 0.006$ & $\simeq -0.884$ \\
$5$ & $\simeq -0.941 \pm 0.011$ & $\simeq -0.955$ \\
$6$ & $\simeq -0.977 \pm 0.004$ & $\simeq -0.983$ \\
\hline\hline
\end{tabular}
\caption{Values of $D_\text{min}$ obtained from the numerical simulations and those using Eq.~(\ref{eq:Dmin}) for $\gamma=3,4,5,6$.}
\label{tab:data_sim}
\end{table}


\end{document}